\documentclass[twocolumn,10pt]{article}

\usepackage{lineno}
\modulolinenumbers[5]

\usepackage{graphicx,setspace}
\usepackage{caption}
\usepackage{epstopdf}
\usepackage{amsmath}
\usepackage{multirow}
\usepackage{times}
\usepackage{amssymb}
\usepackage{dcolumn}%
\usepackage{bm}%
\usepackage{hyperref}
\usepackage{upgreek}
\usepackage{wasysym}
\usepackage{verbatim}
\usepackage[capitalise]{cleveref}
\usepackage[numbers,super,comma,sort&compress]{natbib}

\hypersetup
{
	colorlinks=true,
	linkcolor=blue,
	citecolor=blue
}

 \makeatletter
  \renewcommand\@biblabel[1]{#1.}
  \makeatother

\begin{document}

%\sffamily

\begin{center}
\begin{minipage}{5.5in}

%\noindent {\bfseries \Large Elasticity and yielding in polymeric solids with different kinds of disorder}\\

\noindent {\bfseries \Large Effect of nematic ordering on the elasticity and yielding in disordered polymeric solids}\\

\noindent {Andrea Giuntoli${}^{1}$, Nicola  Calonaci${}^{1, a}$, Sebastiano Bernini${}^{1}$, Dino Leporini${}^{1,2}$\\}

%\vspace{-5mm}

\noindent {${}^{1}$Dipartimento di Fisica ``Enrico Fermi'', 
Universit\`a di Pisa, Largo B.\@Pontecorvo 3, I-56127 Pisa, Italy}

\noindent {${}^{2}$Istituto per i Processi Chimico-Fisici-Consiglio Nazionale delle Ricerche (IPCF-CNR), via G. Moruzzi 1, I-56124 Pisa, Italy\\ 

Correspondence to: D.Leporini (E-mail: dino.leporini@unipi.it)
}

\end{minipage}
\end{center}

%\vspace{-5mm}

\noindent {\em Dated: \today}\\

\begin{center}
\begin{minipage}{5.5in}

{\bf ABSTRACT}
The relation between elasticity and yielding is investigated in a model polymer solid by Molecular-Dynamics simulations.
By changing the bending stiffness of the chain and the bond length, {semicrystalline and disordered glassy polymers} - both with bond disorder - as well as 
{nematic glassy polymers with bond ordering} are obtained.
It is found that in systems with bond disorder the ratio $\tau_Y/G$ between the shear yield strength $\tau_Y$ and the shear modulus $G$ is close to the universal value of the atomic metallic glasses.
The increase of  the local nematic order in glasses leads to the increase of the shear modulus and the decrease of the shear yield strength, 
as observed in experiments on nematic thermosets. A tentative explanation of the subsequent reduction of the ratio $\tau_Y/G$ in terms of  the distributions of the per-monomer stress is offered.

\vspace{1cm}
{\bf Keywords:}  Elasticity, Yield, Polymer Glass, Nematic Glass, Molecular-Dynamics simulation 

\vspace{2cm}

\noindent
${}^{a}$ present address: International School for Advanced Studies (SISSA), via Bonomea 265, 34136 Trieste, Italy

\end{minipage}
\end{center}

\clearpage

\section*{\sffamily \normalsize  INTRODUCTION}
The understanding of the microscopic mechanisms underlying the plastic response of amorphous solids to externally driven deformations
is a current issue in material science research 
both for the lack of a complete theoretical background and its importance in technical applications. \cite{ArgonFractures,StachurskiProgPolymSci97,BarratShearReview,KramerQuestions}
Solids subjected to small deformations respond linearly as expected from elasticity theories. \cite{ArgonPlasticTheory,EshelbyElastic,BudianskyElastic,WardElastic}
An increasing strain on the system causes the increase of internal stress. { Focusing on pure shear deformation,} the elastic modulus $G$ of the system under the studied deformation can be derived
from the slope of the stress-strain curve in the small strain regime \cite{BarratShearReview} 
both locally and globally \cite{BarratAQS13}.
Upon increasing strain, amorphous solids show complex and far from linear behavior \cite{BarratPlastic02,BarratPlastic04,BarratPlastic05},
with heterogeneous and  protocol-dependent \cite{ProcacciaProtocol,ProcacciaCooling} phenomena taking place mainly due to the absence of long-range order \cite{ArgonFractures}.
{ Having reached a characteristic yield strain, corresponding to the shear yield strength $\tau_Y$, the transition from the (reversible) elastic state to the (irreversible) plastic 
one takes place \cite{StachurskiProgPolymSci97,RobbinsHoyJPolymSci06,RottlerSoftMatter10}.
In an ideal elasto-plastic body (Hooke-St.Venant) $\tau_Y$ is the maximum stress \cite{StachurskiProgPolymSci97}.}

{
Despite the complexity of the plastic behavior in amorphous solids at the local scale, some general features have been found in the macroscopic quantities. An interesting aspect of yielding
is that the yield stress is proportional to the elastic modulus. In particular,
for a linear, dislocation-free array of atoms Frenkel derived long time ago the relation   $\tau_Y / G\simeq 1/ ( \pi \sqrt{3} )  \simeq 0.18 \, $ at $T= 0 K$
\cite{FrenkelYield26,FrenkelYieldModel,StachurskiProgPolymSci97}. A more recent experimental work found $\tau_Y/ G  \sim 0.11$ for polymers \cite{ArgonFractures} and the universal value
$0.036 \pm 0.002$  for metallic atomic glasses \cite{JohnsonMetallic}. The ratio $\tau_Y/ G$ depends on the temperature and, for a given temperature, is universal for metallic glasses up to slightly
below the glass transition temperature \cite{JohnsonMetallic}. The finding has been interpreted in terms of similar inter-particle potentials \cite{LernerProcaccia}. 
The microscopic origin of the proportionality between  $\tau_Y$ and $G$ has been rationalized by noting that in both metals and polymers the yield stress is primarily governed by energy storing elastic
processes: dislocation line energy in metals, strain energy around molecular kinks in polymers \cite{ArgonPlasticTheory}. Since the elastic modulus in glassy polymers is dominated by 
inter-molecular forces  \cite{BerniniElasticJPolB15}, it was concluded that the energy barriers to plastic flow in glassy polymers were dominated by intermolecular rather than intra-molecular 
interactions \cite{ArgonFractures}}, so some similarities
can be found in the comparison between atomic and polymeric systems \cite{ArgonSilicon}.
On the other hand, the intra-molecular interactions can have a primary role in determining the structure of a polymer solid upon cooling from the liquid phase, which is of great importance
to determine the elastic properties of the final structure \cite{ProcacciaProtocol, ProcacciaCooling}.

\begin{figure}[t]
\begin{center}
\includegraphics[width= \linewidth]{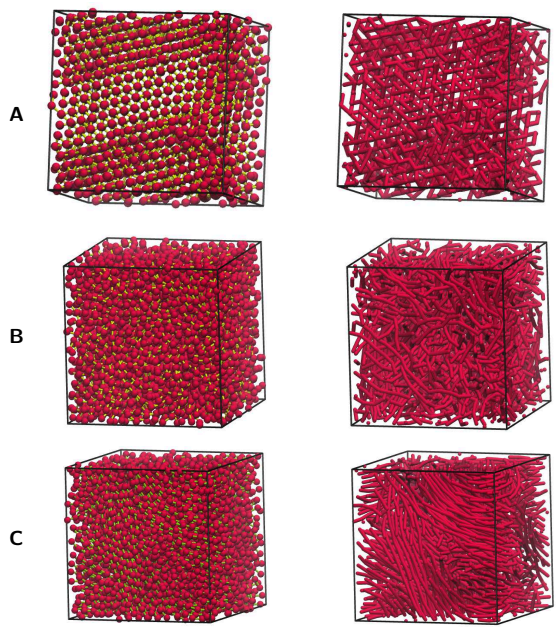}
\end{center}
\caption{{ \small Illustrative snapshots of the different polymer solids under study: {semicrystalline polymer} (A), disordered {glassy polymer} (B), 
nematic {glassy polymer} (C). Monomer position is emphasized in the 
left column, bond orientation is emphasized in the right column. Differently from the {nematic glassy polymer}, 
both {semicrystalline and disordered glassy polymers} exhibit local bond disorder. The snapshots refer to 
chains with bond length $l_b=1.12$ and bending stiffness $k_\theta= 0$ (A), $4$ (B), $12.5$ (C).}}
\label{MDbox} 
\end{figure}

The aim of this work is to investigate the existence of the correlation of shear elastic modulus $G$ and yielding stress $\tau_Y$ in polymer solids by means of molecular dynamics (MD) simulations. 
A model is presented in which the systematic variation of characteristic parameters of the intra-molecular interactions, namely the bond length $l_b$  and the bending stiffness $k_\theta$ of pairs of
contiguous bonds  in a chain, leads to different {semicrystalline, disordered or nematic} structures, see \cref{MDbox}. 
{ For fully-flexible chains with no bending potential ( $k_\theta= 0$ ) we find that the
yield stress increases with the elastic modulus in a way which is very close to the universal law of the atomic metallic glasses \cite{JohnsonMetallic}, suggesting that, in the absence of 
bending stiffness, connectivity and structure play minor roles in the yield process of the present polymer model.}  
Increasing the bending stiffness of the chains causes the increasing growth of the local nematic ordering of near chains.
It is seen that the onset of nematic order {\it increases} the elastic modulus $G$ and {\it  decreases} the yielding stress $\tau_Y$, thus evidencing the different influence of the local order
on the plasticity and the elasticity. A tentative explanation of the subsequent reduction of the ratio $\tau_Y/G$ in terms of  the distributions of the per-monomer stress is offered.

\section*{ \sffamily \normalsize NUMERICAL METHODS}
\label{numerical}
We consider a coarse-grained polymer model of $N_c=160$ linear, unentangled chains with $M=25$ monomers per chain. 
The total number of monomers is $N=4000$. Non-bonded monomers at distance $r$ interact via the truncated and shifted Lennard-Jones (LJ) potential:
\begin{equation}
U^{LJ}(r)=\varepsilon\left [ \left (\frac{\sigma^*}{r}\right)^{12 } - 2\left (\frac{\sigma^*}{r}\right)^6 \right]+U_{cut}
\end{equation}
for $r\leq r_c=2.5\sigma$ and zero otherwise, where $\sigma^{\ast}=2^{1/6}\sigma$, is the position of the potential minimum with depth $\varepsilon$.
The value of the constant $U_{cut}$ is chosen to ensure that $U_{LJ}\left(r\right)$ is continuous at $r=r_c$. Henceforth, all quantities are expressed in terms of reduced units:
lengths in units of $\sigma$, temperatures in units of $\varepsilon/k_B$ (with $k_B$ the Boltzmann constant) and time $\tau_{MD}$ in units of $\sigma \sqrt{m / \varepsilon}$ where 
$m$ is the monomer mass. We set $m = k_B = 1$. The bonding interaction is approximated via the harmonic potential
\begin{equation}
U_{bond}(r)=k_b(r-l_b)^2
\end{equation}
where $l_b$ is the equilibrium bond length and $k_b=300\varepsilon/\sigma^2$ is the bond rigidity. Differently from previous studies  concerning fully-flexible chains \cite{UnivSoftMatter11,PuosiLepoJCPCor12,PuosiLepoJCPCor12_Erratum,Puosi11}, the bending angle interaction between adjacent chemical bonds is included through a potential of the form \cite{KarayiannisBendingAngle15}:
\begin{equation}
U_{bending}=k_\theta(1-\cos\theta_b)
\end{equation}
where $k_\theta$ is the bending stiffness, $\cos\theta_b^i=\vec{b}_{i+1}\cdot\vec{b}_i/||\vec{b}_{i+1}|||\vec{b}_i||$ and the bond vector $\vec{b}_i=\vec{r}_{i+1}-\vec{r}_i$, where $\vec{r}_i$
is the position of the i-th monomer.  
Periodic boundary conditions are used. The study was performed in the $NPT$ ensemble (constant number of particles, pressure and temperature). 
The integration time step is set to $\Delta t=0.005$ time units \cite{Puosi11,UnivPhilMag11}. 
The simulations were carried out using LAMMPS molecular dynamics software (http://lammps.sandia.gov). \cite{PlimptonLAMMPS}

A systematic study is performed by changing the bond length $l_b$ and bending stiffness $k_\theta$. We focus on two families of systems: fully flexible polymers
($k_\theta=0$) with $0.91\leq l_b\leq1.12$, and semi-flexible/stiff polymers with $l_b=1.12$ and $1.0\leq k_\theta\leq12.5$.
All samples are equilibrated in the $NPT$ ensemble at $P=0$. They are initially equilibrated at the following temperatures: $T=1.2$ for $k_\theta < 7$, $T=1.4$ for 
$ 7 \le k_\theta < 12.5$, $T=1.6$ for $k_\theta = 12.5$. Then, they are isobarically cooled down to $T=0$ with a constant quench rate of $|\dot{T}|=2\cdot10^{-6}$. 
Both the equilibration and the quench procedures are close to the one adopted in Ref. \cite{KarayiannisBendingAngle15}.
 Isobaric quenches have also been considered  in other MD investigations of plastic yield in glassy polymers. \cite{RottlerSoftMatter10} 
After the quench, simple shear deformations of the polymer solids at $T=0$, $P=0$ are performed via the Athermal Quasi-Static (AQS) protocol described in details in Ref. \cite{BarratShearReview}.
Initially, the undeformed simulation box containing the sample is a cube with side $L$.
An infinitesimal strain increment $\Delta\varepsilon=10^{-5}  L$ is applied, after which the system is allowed to relax 
in the local potential energy minimum via a suitable minimization algorithm. The procedure is repeated up to the total strain $\varepsilon_{tot}=15 \cdot 10^{-2} L$. 

\begin{figure}[t]
\begin{center}
\includegraphics[width=0.95\linewidth]{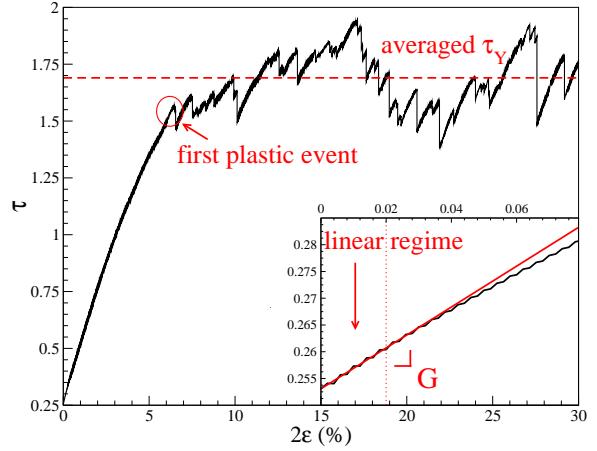}
\end{center}
\caption{ { \small Typical stress-strain curve of our polymer solids under athermal, quasi-static, simple-shear deformation. After a first 'loading' phase, a   plateau-like plastic regime sets
in where a series of sudden stress drops are observed. The yield stress $\tau_Y$ is defined as the average 
value of $\tau$ in the plastic regime  \cite{LernerProcaccia}. The elastic modulus $G$ (see inset) is measured via a linear fit of the stress-strain curve in the linear regime of small deformations
$2\epsilon<0.02$. The plot refers to a system of fully-flexible chains ($k_\theta=0$) with bond length $l_b =1.12$.}}
\label{carico} 
\end{figure}

Simple shear is performed independently in the planes ($xy$, $xz$, $yz$), and at each strain step in the plane {$\alpha\beta$ the corresponding component of the macroscopic stress tensor
$\tau_{\alpha, \beta}$ is taken as the average value of the per-monomer stress  $\tau_{\alpha, \beta}^i$: 
\begin{equation}
\label{stresstensor1}
\tau_{\alpha, \beta} =   \frac{1}{N}  \sum_{i=1}^N  \tau_{\alpha, \beta}^i
\end{equation}
In an athermal system the expression of the per-monomer stress in the atomic representation is \cite{allenMolPhys}: 
\begin{equation}
\label{stresstensor2}
\tau_{\alpha, \beta}^i =  \frac{1}{2 \, v} \sum_{j \ne i} r_{\alpha ij} F_{\beta ij} 
\end{equation}
where $F_{\gamma  k l}$ and $r_{\gamma  k l}$  are the $\gamma$ components of the force between the $k$th and the $l$th monomer and their  separation, respectively, and $v$ is the average per-monomer
volume, i.e. $v = L^3/N$.}
For each plane we then obtain a stress-strain curve, an illustrative example of which is given in \cref{carico}. The result is quite analogous to what reported for many other systems under 
athermal conditions \cite{MottArgonSuter93,FalkViscoPlastic,Maeda2Dplastic,MelandroPlastic,LemaitrePlastic,ProcacciaPiecewise}  with an initial linear increase followed by increasing bending and onset 
of the plastic regime. In particular, similarly to other MD studies of glassy polymers \cite{YieldMDPolymerMM15}, one notices that, in the plastic regime,  the stress levels off to a plateau with 
fluctuations caused by subsequent loading phases and sudden stress drops. We point out that the initial non-zero stress in the unstrained solid seen in  \cref{carico} is a well-known phenomenon 
usually ascribed to the limited size of the simulation cell \cite{HawardCristPhysGlassyPolym}.

We measure the shear elastic modulus $G$ as the slope of the stress-strain curve in the linear regime ($\varepsilon \le 0.01$), via the relation $G  = \tau / 2\varepsilon $, see inset of \cref{carico}.
Following Ref. \cite{LernerProcaccia}, the yield stress $\tau_Y$ is taken as the average value of the stress after the first significant plastic event, defined as the first stress drop 
of at least $\Delta\tau_{th}=0.1$, see \cref{carico}. { This choice is  consistent with other definitions in the presence \cite{RottlerSoftMatter10}, or not \cite{RobbinsHoyJPolymSci06}, 
of strain softening, i.e. the reduction in stress following yield}. The results are robust with respect to changes of $\Delta\tau_{th}$.
Data concerning 16 distinct simulation runs are gathered for each physical state. Each run is averaged over the three planes $xy$, $xz$ and $yz$.

\section*{\sffamily \normalsize RESULTS AND DISCUSSION}
\label{resultsdiscussion}

\subsection*{ \sffamily \normalsize Structural analysis during quench-cooling}
\label{quench}

The elastic properties of amorphous solids strongly depend on the sample preparation \cite{ProcacciaCooling,ProcacciaProtocol}.
Thus, we preliminarily characterize the most relevant structural changes of our systems occurring during the isobaric quench from the liquid  to the athermal solid. 

In order to study more rigorously the structural order of the systems, we resort to the order parameters defined by Steinhardt \textit{et al.} \cite{Steinhardt83}.
One considers in a given coordinate system the polar and azimuthal angles $\theta({\bf r}_{ij})$ and $\phi({\bf r}_{ij})$ of the
vector ${\bf r}_{ij}$ joining the $i$-th central monomer with the $j$-th one belonging to the neighbors within a
preset cutoff distance $r_{cut} = 1.2 \; \sigma^* \simeq 1.35$ \cite{Steinhardt83}. $r_{cut}$ is a convenient definition of
the first coordination shell size \cite{sim}. The vector ${\bf r}_{ij}$ is usually referred to as a ``bond'' and has
not to be confused with the {\it actual} chemical bonds of the polymeric chain.
To define a global measure of the order in the system, one then introduces the quantity:
\begin{equation} \label{Qbarlm_glob}
	\bar{Q}_{lm}^{glob}=\frac{1}{N_{b}}\sum_{i=1}^{N}
\sum_{j=1}^{n_b(i)}Y_{lm}\left[\theta({\bf
r}_{ij}),\phi({\bf r}_{ij})\right]
\end{equation}
where $n_b(i)$ is the number of bonds of $i$-th particle, $N$ is the total number of particles in the system, $Y_{lm}$
denotes a spherical harmonic and $N_b$ is the total number of bonds:
\begin{equation} \label{N_b}
	N_b=\sum_{i=1}^{N} n_b(i) 
\end{equation}
The global orientational order parameter $Q_l^{glob}$ is defined by \cite{Steinhardt83}:
\begin{equation} \label{Ql_glob}
 Q_l^{glob}=\left [ \frac{4\pi}{(2l+1)} \sum_{m=-l}^{l}
|\bar{Q}_{lm}^{glob}|^2 \right ]^{1/2}
\end{equation}
The above quantity is invariant under rotations of the coordinate system and takes characteristic values which can be used to quantify the kind and the degree of
rotational symmetry in the system \cite{Steinhardt83}. In the absence of {\it large}-scale order, the bond orientation is uniformly distributed around the unit sphere and $Q_l^{glob}$ 
is rather small since it vanishes as $ \sim N_b^{-1/2}$ \cite{RintoulTorquatoJCP96}. On the other hand, $Q_6^{glob}$ is very sensitive to any kind of crystallization and increases significantly
when order appears \cite{GervoisGeometrical99,GiuntoliCristallo}.
A local orientational parameter $Q^{loc}_{l}$ can also be defined. We define the auxiliary quantity
\begin{equation} \label{Qbarlm_loc}
	\bar{Q}^{loc}_{lm}( i )=\frac{1}{n_b(i)}\sum_
{j=1}^{n_b(i)}Y_{lm}\left[\theta({\bf r}_{ij}),\phi({\bf
r}_{ij})\right]
\end{equation}
The local order parameter $Q^{loc}_{l}$ is defined as \cite{Steinhardt83}:
\begin{equation} \label{Ql_loc}
 Q^{loc}_{l}=\frac{1}{N} \sum_{i=1}^{N}  \left [
\frac{4\pi}{(2l+1)} \sum_{m=-l}^{l} |\bar{Q}^{loc}_{lm}( i
)|^2 \right ]^{1/2}
\end{equation}
In general $Q^{loc}_{l}\ge Q_l^{glob}$. In the presence of ideal order, {\it all} the particles have the {\it same} neighborhood configuration, and the equality $Q^{loc}_{l} = Q_l^{glob}$
follows.

\begin{figure}[t]
\begin{center}
\includegraphics[width=  \linewidth]{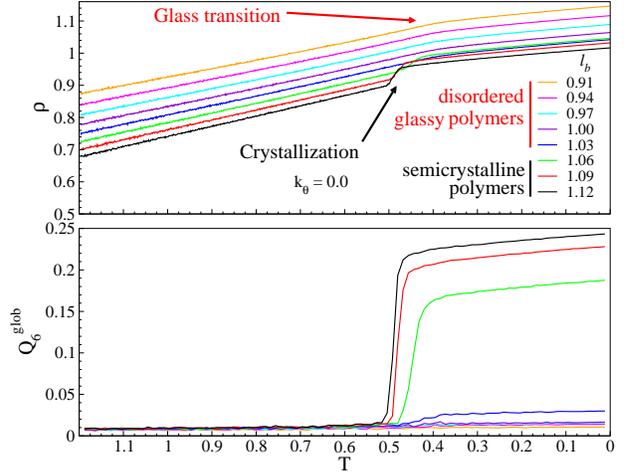}
\end{center}
\caption{ { \small Density $\rho$ (upper panel) and global order parameter $Q^{glob}_6$ (lower panel) of a melt of fully-flexible chains ($k_\theta=0$) with different bond length
$l_b$ during the isobaric quench 
from the liquid to the solid phase. Chains with short bond length ($l_b\leq1.03$) form {disordered glassy polymers} since the bond length is  incommensurate with the 
Lennard-Jones length scale $\sigma^\ast \simeq 1.12$.
Chains with bond length comparable to $\sigma^\ast \simeq 1.12$ exhibit steep increase of the density $\rho$ and the global order parameter $Q^{glob}_6$ upon cooling.
 In the latter case, the local-order analysis presented in Fig.\ref{corrq6q4} clarifies that the corresponding solids at $T=0$ are {semicrystalline polymers}
with coexisting polymorphs.}}
\label{BNDdensq6} 
\end{figure}

We first examine the density and the global order of fully-flexible chains ($k_\theta=0$). 
The global positional order of the monomers is monitored via the Steinhardt order parameter $Q^{glob}_6$. \cref{BNDdensq6} plots the increase of both the density $\rho$ and
the order parameter $Q^{glob}_6$ for different bond lengths $l_b$ while decreasing the temperature at constant pressure $P=0$ from the initial liquid state to the final solid state.
Fully flexible polymers either exhibit global order or glassify upon cooling, depending on the bond length $l_b$.
Global order is revealed by sharp jumps in density $\rho$ and $Q^{glob}_6$ for $l_b=1.06,1.09,1.12$.
A local-order analysis, presented later in the paper, clarifies that the states with global order are {semicrystalline polymers} with coexisting polymorphs at $T=0$.
Systems with shorter bond length form {glassy polymers}, with no significant global order.  If the bond length is comparable to the monomer size, $l_b \approx \sigma^\ast \simeq 1.12$,
the formation of ordered structures is to be expected \cite{SimmonsBcc13,KarayiannisCEE09,KarayiannisPRL09HSC,KarayiannisSM10AthHSC,GiuntoliCristallo}, whereas shorter bonds are known
\cite{Hamley2007,DoyeFrustratCrystalPhysChemChemPhys07,MilsteinPRB70,SimmonsBcc13}
to cause geometrical frustration which hinders the crystallization process, thus favoring the formation of {disordered glassy polymers}.

We now turn to semi-flexible and stiff chains ($k_\theta > 0$). Since the reduced flexibility favors local nematic ordering,
i.e. the alignment of near bonds, we divide the sample in $n^3$ cells with side $L/n$ and  define the bond-orientation order parameter in the $i$-th cell as \cite{LuoSommerLocalNematicOrderM11}
\begin{equation}
{S_i}=\sqrt{\frac{3}{2}Tr(q_i^2)}, \quad q_{i,\alpha\beta}=\langle\hat{b}_{\alpha}\hat{b}_{\beta}-\frac{1}{3}\delta_{\alpha\beta}\rangle_i
\label{orderparameter0}
\end{equation}
where $1\le i\le n^3$, $Tr$ is the trace operator,  $q_i$ is a $3 \times 3$ orientational tensor with components of $q_{i,\alpha\beta}$,
$\hat{b}_{\alpha}$ and $\hat{b}_{\beta}$ are the Cartesian components of the normalized bond vectors $\vec{b}$ and the statistical average $\langle...\rangle_i$ is performed
on all the bonds of the $i$-th cell. {Following Karayiannis and coworkers \cite{KarayiannisBendingAngle15}, 
we initially choose $n=6$ corresponding to cells with side of about $2-3$ monomer diameters.
An average local bond-orientation order parameter is then defined as \cite{LuoSommerLocalNematicOrderM11}
\begin{equation}
S_{loc}^{bond}= \frac{1}{216} \sum_{i=1}^{216} S_{i}
\label{orderparameter}
\end{equation}
}
\begin{figure}[t]
\begin{center}
\includegraphics[ width= 0.85 \linewidth]{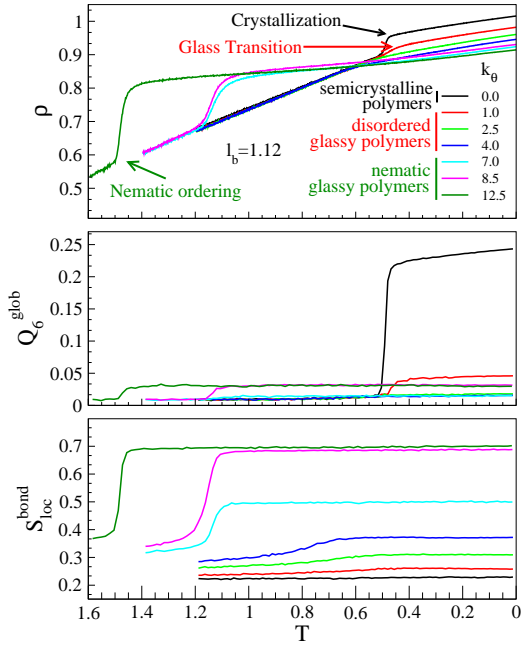}
\end{center}
\caption{{ \small Density $\rho$ (top), global order parameter $Q^{glob}_6$ (middle) and local bond-orientation order parameter {$S_{loc}^{bond}$} 
(bottom) of a melt of chains with increasing 
bending stiffness during the isobaric quench from the liquid to the solid phase. Bond length $l_b=1.12$. For  fully-flexible chains ($k_\theta=0$) a steep increase of the density $\rho$ and the global 
order parameter $Q^{glob}_6$ is revealed at $T\simeq 0.5$. For semi rigid/stiff chains ($k_\theta>0$): i) all the final solid states are {glassy polymers} ($Q^{glob}_6 < 0.05$), ii) 
on cooling, the bending stiffness triggers a transition to a nematic state with bond ordering occurring in the liquid phase and freezing below the glass transition.}}
\label{STFdensS}
\end{figure}
\begin{figure}[t]
\begin{center}
\includegraphics[ width=   \linewidth]{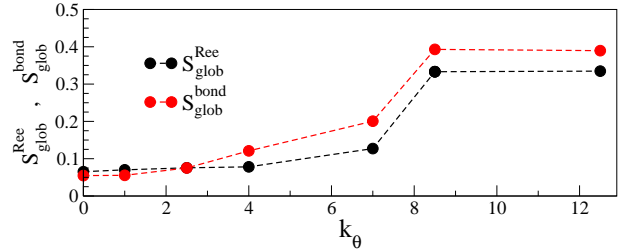}
\end{center}
\caption{{ \small Dependence of the {global} bond-orientation order parameter {$S_{glob}^{bond}$} and the global {chain-orientation} order parameter 
{$S_{glob}^{Ree}$} on the bending stiffness by increasing $k_\theta$ at $T=0$. 
Bond length $l_b=1.12$. Note that, even for high bending stiffness, the global order is not strong despite the local ordering shown in Fig.\ref{STFdensS} (bottom). See Fig.\ref{MDbox} for selected snapshots.}}
\label{S1Sg}
\end{figure}

{The $S_{loc}^{bond}$ order parameter  ranges between $S_{loc}^{bond}=1$ (perfect alignment) and $S_{loc}^{bond}=0$ (random orientation).} 
\cref{STFdensS} plots the density $\rho$, the order parameter $Q^{glob}_6$ and 
the local bond-orientation order parameter {$S_{loc}^{bond}$} of systems with bond length $l_b=1.12$ and different bending stiffness  $k_{\theta}$, during the isobaric quench
from the initial liquid state to the final solid state. 
The latter exhibits global order only if the chains are fully flexible ($k_{\theta}=0$), as signaled by the jumps of both the density and the global order 
parameter at $T\simeq 0.5$, otherwise {glassy polymers} with small global order ($Q^{glob}_6 < 0.05$) are obtained. It is seen that the increasing bending stiffness of the chains 
triggers a transition to a nematic state with considerable local alignment of the bonds, as detected by the increase of the bond-orientational order parameter {$S_{loc}^{bond}$}. 
The resulting local orientational order freezes below the glass transition, yielding a {nematic glassy polymer}. 

It is interesting to consider the {global} bond-orientation order. To this aim, we set $n=1$ and define the {global} bond-orientation order parameter $S_{glob}^{bond}$ as $S_1$ from eq.\ref{orderparameter0} to perform the average of the bond orientation over a {\it single} cell coinciding with all the sample. 
The quantity is plotted in Fig.\ref{S1Sg}. On increasing the  bending stiffness $k_\theta$ at $T=0$, {$S_{glob}^{bond}$} starts from $\sim 0.05$ for fully-flexible chains ($k_\theta=0$), 
then increases and levels off at the plateau level $S_{glob}^{bond} \simeq 0.38$ for  $k_\theta \gtrsim 8.5$. This suggests that the sample is {\it locally oriented} 
(high {$S_{loc}^{bond}$}), but {\it macroscopically nearly isotropic} (small {$S_{glob}^{bond}$}) for strong bending stiffness.
To corroborate the previous conclusion, we consider the alignment of the end-to-end unit vector of the chains via the global {chain-orientation} order parameter 
{$S_{glob}^{Ree}$} \cite{KarayiannisBendingAngle15}. 
By construction, {$S_{glob}^{Ree}$} spans the range between {$S_{glob}^{Ree} = 1$} (perfect alignment of all the chains) and 
{$S_{glob}^{Ree} = 0$} (random orientation). Fig.\ref{S1Sg} shows that {$S_{glob}^{Ree}$} increases with the bending stiffness but it is not large. 
\begin{figure}[t]
\begin{center}
\includegraphics[width= \linewidth]{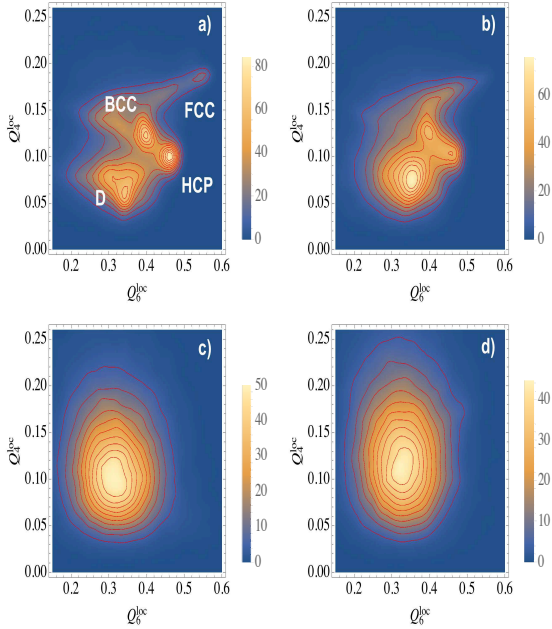}
\end{center}
\caption{{ \small Bivariate distributions of the local order parameters $Q^{loc}_{4}$ and $Q^{loc}_{6}$ for characteristic states at $T=0$:  (a) {semicrystalline polymer} with 
$l_b=1.12$ and $k_\theta=0$; 
(b) {semicrystalline polymer} with $l_b=1.06$ and $k_\theta=0$; (c) disordered {glassy polymer} with $l_b=1.12$ and $k_\theta=4$; 
(d) nematic {glassy polymer} with $l_b=1.12$ and $k_\theta=12.5$. In panel a) the regions corresponding 
to the BCC, FCC and HCP structures at the level of the first neighbor shell are marked. The region "D" labels states with first neighbor shell different from the BCC, FCC and HCP ones. 
The contour lines have equal contour interval and divide the whole elevation range evenly.
}}
\label{corrq6q4}
\end{figure}

In order to gain more insight into the structure of the polymeric solids Fig.\ref{corrq6q4} presents the correlation plots of the local order parameters $Q^{loc}_{4}$ and $Q^{loc}_{6}$, 
characterizing the order of the {\it first} neighbor shell of each monomer. Fig.\ref{corrq6q4}a shows the complex nature of the solid state with $l_b=1.12$ and $k_\theta = 0$, corresponding to 
fully-flexible chains with bond length comparable to the monomer size.
Four different regions with highly-correlated pairs
$(Q^{loc}_{4},Q^{loc}_{6})$  are apparent. According to previous studies \cite{GiuntoliCristallo,LocalOrderJCP13}, two of them signal face-centered cubic (FCC) and hexagonal close packed (HCP)
local packings. For the same polymer model with $k_\theta = 0$ and $l_b=1.12$ at  $T=0$, FCC and HCP close packed structures together with other (unspecified) non close-packed environments were detected 
\cite{KarayiannisBendingAngle15}. 
We also identify high correlations in the region $(Q^{loc}_{4},Q^{loc}_{6}) \simeq (0.12,0.4)$. These values are ascribed to a deformed body-centered cubic (BCC) structure 
with $(Q^{loc}_{4},Q^{loc}_{6})$ pair significantly different from the ideal BCC { due to poor stability of the BCC lattice \cite{GiuntoliCristallo,MisraCrystStability1940}}. 
On the basis of previous studies \cite{FrenkelLJCrystalJCP96}, we believe that such BCC structures were nucleated as metastable regions during the quench and frozen in the solid phase at $T=0$. 
BCC structures have been reported for the present model with $k_\theta = 0$ and $l_b \simeq 0.97$ in the crystallization of a polymer melt exposed to well-ordered walls \cite{SimmonsBcc13} and in 
the spontaneous isothermal crystallization of an unbounded polymer melt \cite{GiuntoliCristallo}.  { The D region in Fig.\ref{corrq6q4}a represents environments with first neighbor shell different 
from the BCC, FCC and HCP ones. }

In summary, the solid state of fully-flexible chains with bond length comparable to the monomer size, $l_b =1.12  \approx \sigma^\ast$,  is semicrystalline with coexisting polymorphs. 
The structure of the solid appears to be much less heterogeneous by decreasing the bond length or increasing the bending stiffness. In fact, Fig.\ref{corrq6q4}b shows that, 
if $l_b = 1.06$ with $k_\theta = 0$,  the D region is enhanced to the detriment of the BCC, FCC and HCP regions. For  $l_b \le 1.03$ and $k_\theta = 0$ the solid is a disordered glass and only the 
D region is apparent (not shown). A similar finding is observed by keeping $l_b =1.12$ and increasing the strength of the bending potential, see c) and d) panels of Fig.\ref{corrq6q4}. 
{ Then, we see that the D region is characteristic of our glassy systems}.

We note that Fig.\ref{corrq6q4}d shows two weak lobes located at $Q^{loc}_{6} \simeq 0.48$ with $Q^{loc}_{4} \simeq 0.09$ and $0.175$. By comparison with panels a) and c) of Fig.\ref{corrq6q4}, 
the finding suggests 
reentrant FCC and HCP ordering on increasing the strength of the bending potential with $l_b =1.12$. The finding is consistent with the results reported by Karayiannis and coworkers 
\cite{KarayiannisBendingAngle15} where the fraction of sites with close-packed order (FCC or HCP similarities) is close to one in systems with 
{$S_{glob}^{Ree} \simeq 1$}, i.e. nearly straight chains, 
and high local orientation order, {$S_{loc}^{bond} \sim 0.95$}. We remind that in our case {$S_{glob}^{Ree}$} and {$S_{loc}^{bond}$} are not larger than about $0.33$ and $0.7$, 
respectively. Incidentally, the fact that we find less 
global and local orientational order with the same polymer model with respect to Ref.\cite{KarayiannisBendingAngle15}  is ascribed to the smaller size of our sample.

\begin{figure}[t]
\begin{center}
\includegraphics[width=0.99 \linewidth]{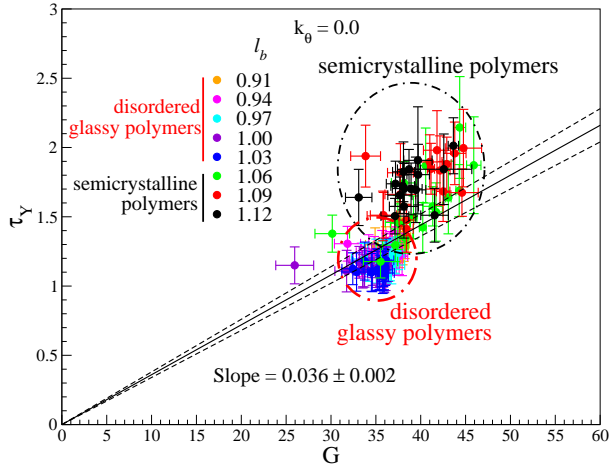}
\end{center}
\caption{ { \small Correlation plot of  the elastic modulus $G$ and the average yield stress $\tau_Y$ for the athermal solids made by fully-flexible chains with different bond lengths. 
The error bars of $\tau_Y$ and $G$ reflect the fluctuation of the stress during the steady state of the plastic regime, and the uncertainty of the fit in the linear elastic regime, see \cref{carico}, 
respectively.  {Semicrystalline polymers} and disordered glassy polymers exhibit correlations in two different regions, inside of which the influence of the bond length is minor.  
{The black continuous line is the 
universal law of metallic glasses $\tau_Y = m \, G$ with slope $m = 0.036 \pm 0.002$ \cite{JohnsonMetallic}. The uncertainty on the $m$ parameter is bounded by the two dashed lines.}}
}
\label{BNDcorr}
\end{figure}

\subsection*{\sffamily \normalsize Correlation between yield stress and shear modulus }
\label{shear}
\cref{BNDcorr} is a correlation plot of the average yield stress $\tau_Y$ and the elastic shear modulus $G$  for the solids made by fully-flexible chains with different bond lengths $l_b$. 
The plot presents the data on a run-by-run basis, i.e. no average between runs with the same bond length is performed. 
A general tendency of the yield stress $\tau_Y$ to increase with the modulus is observed.
It is seen that {disordered glassy polymers} exhibit limited changes of both $\tau_Y$ and $G$, whereas
 {semicrystalline polymers} show a wider distribution across the different runs. We ascribe the effect to the polymorphic character of the ordered solids \cite{GiuntoliCristallo}.
Also, {semicrystalline polymers} show higher $G$ and $\tau_Y$ values with respect to {disordered glassy polymers}, meaning that the increased order of the monomeric 
arrangement causes the system to react to shear deformations with stronger internal stresses with respect to its amorphous counterpart both in the linear regime and at the yield point. 
In \cref{BNDcorr} we superimpose to our data the characteristic universal law of the metallic {\it atomic} glasses, i.e. the line $\tau_Y = m \, G$ with  $m = 0.036 \pm 0.002$ \cite{JohnsonMetallic}. 
Deviations are apparent but not large, thus suggesting that, in the absence of bending stiffness, connectivity and structure play minor roles in the yield process of the present polymer model. 

\begin{figure}[t]
\begin{center}
\includegraphics[width=0.99 \linewidth]{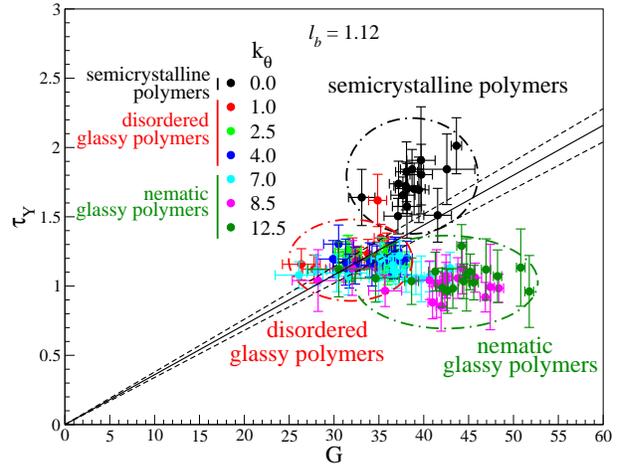}
\end{center}
\caption{{\small
Correlation plot of  the elastic modulus $G$ and the average yield stress $\tau_Y$ for  the athermal solids made by chains with different bending stiffness. 
Bond length $l_b=1.12$. As in \cref{BNDcorr} the black continuous line is the universal law of metallic glasses $\tau_Y = m \, G$ with slope $m = 0.036 \pm 0.002$ \cite{JohnsonMetallic}. 
Differently from {semicrystalline and disordered glassy polymers}, {\it {nematic glassy polymers}} exhibit large deviations from that law.}}
\label{STFcorr}
\end{figure}

The introduction of bending stiffness, $k_{\theta}\neq0$, and the subsequent nematic order provide a different scenario. This is clearly visible in the correlation plot of the average yield stress 
$\tau_Y$ and the elastic shear modulus $G$, see \cref{STFcorr}. For low and intermediate bending stiffness, $k_{\theta}\le 4$, the solids are {semicrystalline polymers} 
or microscopically disordered {glassy polymers} 
respectively, with ratio  $\tau_Y / \, G$  close to the characteristic universal value $0.036 \pm 0.002$ of the atomic metallic glasses \cite{JohnsonMetallic}, as in \cref{BNDcorr}. 
For {nematic glassy polymers}, $k_{\theta}\ge 7$, the ratio $\tau_Y / \, G$ decreases by increasing the bending stiffness of the chain.

\begin{figure}[t]
\begin{center}
\includegraphics[width=0.99\linewidth]{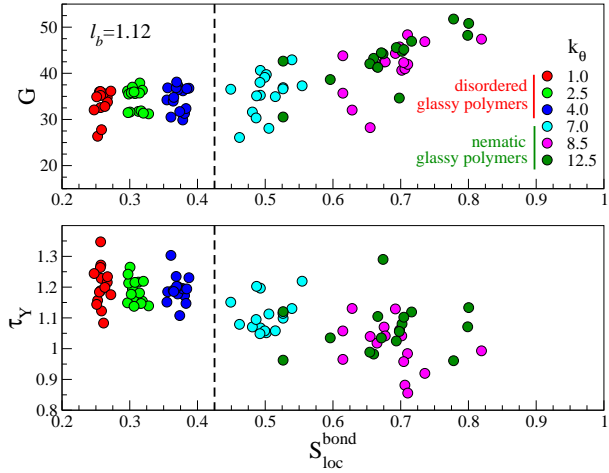}
\end{center}
\caption{Correlation between the elastic modulus (top) and the yield stress $\tau_Y$ (bottom) with the  local bond-orientation order parameter {$S_{loc}^{bond}$,
Eq.\ref{orderparameter}.} 
Bond length $l_b=1.12$. The dashed line divides the regions pertaining to disordered (left) and nematic (right) {glassy polymers}. 
It is seen that in {nematic glassy polymers} the increasing local bond alignment {\it increases} the elastic modulus and {\it decreases} the yield stress.}
\label{decoupling}
\end{figure}
\begin{figure}[t]
\begin{center}
\includegraphics[width=0.99\linewidth]{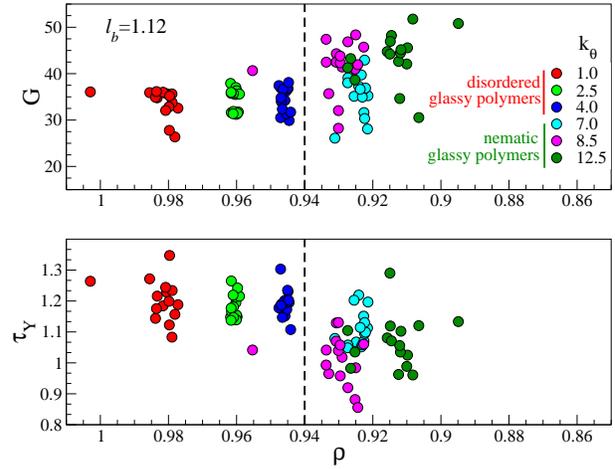}
\end{center}
\caption{Correlation between the elastic modulus (top) and the yield stress $\tau_Y$ (bottom) with the density. Bond length $l_b=1.12$. 
The dashed line divides the regions pertaining to disordered (left) and nematic (right) {glassy polymers}. 
Notice that the {nematic glassy polymers} with $k_\theta = 7$ and $k_\theta = 8.5$ have 
nearly identical densities but rather different local nematic order, see \cref{STFdensS} top. The fact that their modulus and yield stress are appreciably different 
signals the influence of the bond ordering.}
\label{decoupling2}
\end{figure}

We have investigated the origin of the deviations of the ratio $\tau_Y / \, G$ from the characteristic universal value of the atomic metallic glasses.  Elasticity and yielding of polymeric solids 
are both affected by density \cite{StachurskiProgPolymSci97} and local nematic order  
\cite{DouglasThermoset02,DouglasNematicThermosetMM98,OrtizNematic}, two properties which are changed  by varying the  bending stiffness, see \cref{STFdensS}. 
We first consider the influence of nematic order.  \cref{decoupling} shows that in {nematic glassy polymers}, on increasing the local orientational  order of the bonds, 
the elastic modulus {\it increases} and the yield stress {\it decreases}. 
A similar effect has been observed by Ortiz et al \cite{OrtizNematic} in the glassy phase of a macroscopically disordered, liquid-crystalline thermoset, 
where changing the densely cross-linked network structure from an ensemble of randomly oriented rigid-rods to local nematic increases the modulus and decreases the yield stress, 
see Table 3 and 4 of Ref. \cite{OrtizNematic}.  Since the increase of the nematic order is accompanied by the decrease of the density (apart from a small inversion on increasing $k_{\theta}$ from
7 to 8.5, see \cref{STFdensS} top), we have also examined the role of the density. \cref{decoupling2}
shows that in disordered {glassy polymers}, in spite of a density change of about $6 \%$ neither $G$ nor $\tau_Y$ change appreciably. 
Changes are visible in {nematic glassy polymers} where density changes are smaller due to the better packing. This suggests that density plays a minor role, with respect to nematic order, 
in setting both the modulus and the yield stress. In this regard, the comparison between the {nematic glassy polymers} with bending stiffness $k_{\theta}=7$ and $k_{\theta}= 8.5$ 
provides more insight. 
The two systems have rather comparable density but quite different local nematic order, see \cref{STFdensS}. \cref{decoupling2} shows that their moduli (yield stress) are distinctly different, 
increasing (decreasing) with the local nematic order. All in all, the discussion of \cref{decoupling} and \cref{decoupling2} points to the conclusion that in the  polymer model under study elasticity 
and yielding are more affected by the local nematic order than packing. The weak role of packing was also noted in other studies concerning the fast dynamics of polymers  \cite{BarbieriGoriniPRE04}.

\begin{figure}[t]
\begin{center}
\includegraphics[width=0.99\linewidth]{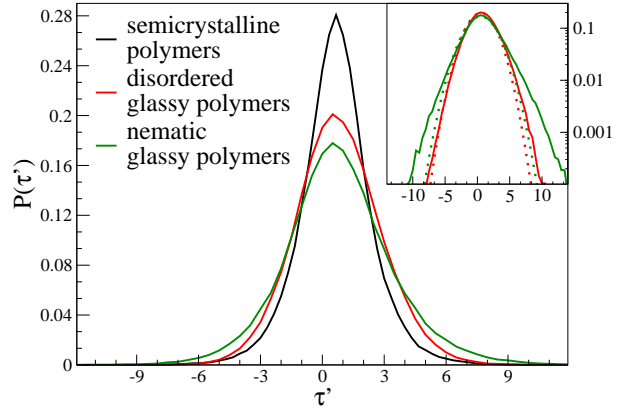}
\end{center}
\caption{Distribution of the { per-monomer} shear stress, Eq.\ref{stresstensor2}, in {semicrystalline polymers} ($k_\theta = 0$), disordered ($k_\theta = 4.0$) and nematic 
($k_\theta = 12.5$) {glassy polymers}. 
All systems have  bond length $l_b=1.12$ and subjected to average stress $\tau = 0.5$, exceeding the linear elastic regime but still far from the region where the sharp plastic drops are observed, 
see \cref{carico}. Inset: comparison between the distributions of the disordered  and the {nematic glassy polymers}.
The dotted curves are the best-fit with gaussians showing that the distribution of the nematic {glassy polymer} exhibits heavy non-gaussian tails.}
\label{distrib}
\end{figure}

{Finally, \cref{distrib} plots the { per-monomer} shear stress distributions  in {semicrystalline polymers} ($k_\theta = 0$), disordered ($k_\theta = 4.0$) and 
nematic ($k_\theta = 12.5$) {glassy polymers}. 
It is seen that the nematic {glassy polymer} exhibits the broadest distribution with heavy non-gaussian tails. This finding suggests a tentative explanation 
of the reduction of the ratio $\tau_Y/G$ in {nematic glassy polymers} with respect to {semicrystalline polymers} and 
disordered {glassy polymers}, see \cref{STFcorr}. 
In fact, it is known that application of a local stress $\tau'$ decreases the energy barrier $\Delta E$ for plastic rearrangements to $\Delta E - \tau' V^\star$ where $V^\star$ is 
an activation volume \cite{StachurskiProgPolymSci97,ArgonPlasticTheory}. If the energy barrier is due to the elastic resistance of the surroundings  treated as an isotropic continuum, 
one finds  $\Delta E = G V^\dagger$, where $V^\dagger$ is a further activation volume distinct from $V^\star$  \cite{Frenkel,StachurskiProgPolymSci97,LiGilmanElsticDisclinJAppPhys70,ArgonFractures}. 
If one assumes that yielding at $T= 0K$ occurs when the energy barrier vanishes, one finds that the local yield  stress $\tau'_Y = V^\dagger/ V^\star G$. If the stress distribution is narrow, 
the local stresses little differ from the average stress,  $\tau_Y \simeq \tau'_Y = V^\dagger/ V^\star G$, and one recovers the usual coupling between the elastic modulus and the macroscopic yield 
stress. Otherwise, if the distribution broadens, highly stressed regions yield when the average stress is much less than their stress $\tau'$, so that $\tau_Y < V^\dagger/ V^\star G$, namely the ratio 
$\tau_Y/ G$ decreases with respect to the characteristic value for systems with narrow stress distribution.  We are aware that our arguments are rather rough. Nonetheless, they offer a consistent 
picture leading to the scenario of \cref{STFcorr}.}

\section*{\sffamily \normalsize CONCLUSIONS}
\label{conclusions}
Elasticity and yielding in polymer solids have been investigated by MD simulations of a coarse-grained model of linear chains with different bending stiffness and bond length. Following the isobaric quench at $T=0$, three kind of distinct structures are observed: 
\begin{itemize}
\item {disordered glassy polymers}: systems with no positional order and and no bond-orientational order,
\item {nematic glassy polymers}: systems with no local positional order but a strong degree of local bond ordering,
\item semicrystalline polymers: systems with local positional order and no bond-orientational order.
\end{itemize}
Note that in this model system semicrystalline polymers do not have any bond orientational order but in other models, e.g. the CG-PVA model \cite{MeyerMullerPlatheMM2001}, short chains with $M \le 30$ form unfolded semicrystalline structures with both local positional (2D hexagonal) and local bond-orientational order \cite{MeyerMullerPlatheJCP2001,MeyerFold02}.

Under simple shear deformations, it is found that in systems with bond disorder the ratio $\tau_Y/G$ between the 
shear yield strength $\tau_Y$ and the shear modulus $G$ is close to the universal value of the atomic metallic glasses. In the presence of increasing nematic ordering the shear modulus of the 
{glassy polymer}
increases while the shear yield strength decreases, thus reducing the ratio $\tau_Y/G$. The finding parallels similar experimental results concerning nematic thermosets. The results suggest that 
nematic order has stronger influence than density on elasticity and yielding. { A tentative explanation of the reduction of the ratio $\tau_Y/G$ in {nematic glassy polymers} with 
respect to {semicrystalline polymers} and disordered {glassy polymers} is offered, pointing out the larger width of the { per-monomer} stress distributions.}

\subsection*{\sffamily \normalsize ACKNOWLEDGMENTS}

F. Puosi is gratefully thanked for helpful discussions. A generous grant of computing time from IT Center, University of Pisa and Dell EMC${}^\circledR$ Italia is gratefully acknowledged.

%\sffamily
%\bibliography{biblio.bib}

\providecommand*{\mcitethebibliography}{\thebibliography}
\csname @ifundefined\endcsname{endmcitethebibliography}
{\let\endmcitethebibliography\endthebibliography}{}

\end{document}